\documentclass[12pt]{iopart}
\input{psfig.sty}

\usepackage{color}

\newcommand{\prd}{{Phys. Rev. D}}

\newcommand{\be}{\begin{equation}}
\newcommand{\ee}{\end{equation}}
\newcommand{\bea}{\begin{eqnarray}}
\newcommand{\eea}{\end{eqnarray}}

\begin{document}


\title[Dark Energy Evolution]{Narrowing Constraints with Type Ia Supernovae: Converging on a Cosmological Constant}


\author{Scott Sullivan$^1$, Asantha Cooray$^1$, Daniel E. Holz$^{2,3}$}
\address{$^1$Center for Cosmology, Department of Physics \& Astronomy, University of California, Irvine, 92697\\
$^2$Theoretical Division, Los Alamos National Laboratory,
Los Alamos, NM 87545\\
$^3$ Department of Astronomy \& Astrophysics,
University of Chicago, Chicago, IL 60637}

\begin{abstract}
We apply a parameterization-independent approach to fitting the dark energy
equation-of-state (EOS). Utilizing the latest type Ia supernova data, combined
with results from the cosmic microwave background and baryon acoustic
oscillations, we find that the dark energy is consistent with a cosmological
constant.  We establish independent estimates of the evolution of the dark
energy EOS by diagonalizing the covariance matrix. We find three independent
constraints, which taken together imply that the equation of state is more
negative than -0.2 at the 68\% confidence level in the redshift range $0<z<1.8$,
independent of the flat universe assumption.  Our estimates argue against
previous claims of dark energy ``metamorphosis,'' where the EOS was found to be
strongly varying at low redshifts. Our results are inconsistent with extreme
models of dynamical dark energy, both in the form of ``freezing'' models where
the dark energy EOS begins with a value greater than -0.2 at $z > 1.2$ and rolls
to a value of -1 today, and ``thawing'' models where the EOS is near -1 at high
redshifts, but rapidly evolves to values greater than -0.85 at $z < 0.2$.
Finally, we propose a parameterization-independent figure-of-merit, to help
assess the ability of future surveys to constrain dark energy. While previous
figures-of-merit presume specific dark energy parameterizations, we suggest a
binning approach to evaluate dark energy constraints with a minimum
number of assumptions.
\end{abstract}

\maketitle

\section{Introduction}
\label{sec:intro}

Distance estimates to Type Ia supernovae (SNe) are currently a preferred probe of the expansion
history of the Universe \cite{Rieetal04}, and have led to
the discovery that the expansion is accelerating
\cite{perlmutter-1999,Rie06,Wood,Davis07}. It is now believed that a mysterious dark energy component,
with an energy density $\sim$70\% of the total
energy density of the universe, is responsible for the accelerated expansion \cite{Spergel,Spergel06}.
While the presence of acceleration is now well established by various cosmological probes,
the underlying physics remains a complete mystery \cite{Padmanabhan}. As the precise nature of the dark
energy has profound implications, understanding its properties is one of the
biggest challenges of modern physics.

With the advent of large surveys for Type Ia supernovae, such as the Supernova
Legacy Survey (SNLS)\footnote{http://www.cfht.hawaii.edu/SNLS/} \cite{Astier}
and Essence\footnote{http://www.ctio.noao.edu/\~wsne/} \cite{Wood}, among
others, it is now possible to consider detailed studies of the expansion history
of the Universe, and shed light on the underlying physics responsible for the
acceleration.
Although the dark energy may be complex, thus far it is generally described by a
cosmological constant, or through a simple dynamical
component such as a single scalar field rolling down a potential
\cite{Caldwell98}.
The observational data is then used to constrain these simple models, generally
in the from of determining a dark energy equation-of-state (EOS) describing the
ratio of its pressure to its density \cite{Garnavich}, or by measuring dynamical
parameters such as the cosmic jerk \cite{Visser}.  Using the EOS as the primary
variable, several studies have considered how current and future data might be
used to make statements on the physics responsible for dark energy \cite{HT99},
including attempts to establish the shape of the scalar field potential
\cite{Sahlen,Li07}.

When model fitting data it is generally assumed that the dark energy EOS as a
function of redshift, $w(z)$, follows a certain predetermined evolutionary
history. Common parameterizations include a linear variations with redshift,
$w(z)=w_0+w_z z$ \cite{CooHut99}, an evolution that asymptotes to a constant $w$
at high redshift, $w(a)=w_0 + w_a (1-a)$ with $a$ as the scale factor
\cite{Lin03}, or an evolution with an EOS of the form $w(z)=w_0-\alpha \ln(1+z)$
\cite{Gerke}. Unfortunately, fitting data to an assumed functional form leads
to possible biases in statements one makes about the dark energy and 
its evolution, especially if the true behavior of the dark energy EOS differs
significantly from the assumed functional form \cite{Li07}. Moreover, statements
related to the dark energy EOS are often made under the assumption of a
spatially flat universe, while there still exists percent-level uncertainties on
the curvature.

Instead of using a parameterized form for $w(z)$, one can also utilize a variant
of the principal component analysis advocated in Ref.~\cite{HutSta03} to
establish the EOS with redshift. This was first applied in Ref.~\cite{HutCoo} to
a set of supernova data from Ref.~\cite{Rieetal04}. Recently, Riess et
al. \cite{Rie06} analyzed a new set of $z > 1$ SNe from the Hubble Space
Telescope combined with low redshift SNe, and by making use of the same
technique of decorrelating the $w_i$ binned estimates, established that the EOS
is not strongly evolving.  Since the analysis of Riess et al. \cite{Rie06}, the
supernovae sample size has increased by at least by a factor of 2 through the
Essence survey \cite{Wood}. The SNe light curves from several independent
datasets have been analyzed with a common method to extract distance moduli in
Ref.~\cite{Davis07}, and we use this publicly available
data\footnote{http://www.ctio.noao.edu/wproject/wresults/} to extract the EOS.

We combine the different distance measurements to extract independent estimates
of the EOS when binned in several redshift bins in the redshift range
$0<z<1.8$. We make use of these binned and uncorrelated estimates of the EOS to address a
simple question: is the dark energy consistent with a cosmological constant?
For a cosmological constant $w(z)=-1$ exactly, while
dynamical dark energy models lead to an EOS that either evolves from a large
value to -1 today (models which are categorized as ``freezing'' in
Ref.~\cite{Caldwell}) or from -1 at high redshift to large values today
(``thawing'' models of Ref.~\cite{Caldwell}). In our analysis we also allow
departures from a flat universe, but allow curvature to be constrained based on
complementary information from the cosmic microwave background ({\it WMAP};
\cite{Spergel06}) and baryon acoustic oscillation distance scales \cite{Eis05}.
As discussed in Ref.~\cite{HutCoo}, our analysis is facilitated by the fact that
our measurements of the EOS are completely uncorrelated.  Although we focus on the
EOS, we note that one can also extract uncorrelated estimates of other
parameters related to the expansion history and dark energy, such as the dark
energy density \cite{Wang}.  However, unlike the case with the EOS, such estimates
cannot be used to directly address the simple question of whether the dark energy is a
cosmological constant.

We make use of our binned estimates to comment on several new developments
related to dark energy studies.  First, in addition to addressing the extent
to which $w(z)=-1$ in present data, we also comment on the extent to which
dynamical dark energy models, such as ``freezing'' and ``thawing'' models
\cite{Caldwell}, may be distinguished or ruled out with current cdistance data.
We also discuss the role uncorrelated, independent EOS measurements can play in
furthering our understanding of the dark energy. Recently, the Dark Energy Task
Force (DETF) \cite{DETF} has suggested a figure-of-merit to compare the
abilities of different experiments to extract information related to dark
energy. This is done in terms of the inverse area of the error ellipse of the
equation of state and its evolution with redshift, utilizing either the Linder
parameterization \cite{Lin03} or with two parameters of the form
$w(a)=w_p+(a_p-a)w_a$ \cite{HT99}.  The discussion is then restricted to a two
parameter description of the dark energy equation of state, assuming a very
specific evolutionary behavior. The ability to extract information about dark
energy from current and future experiments thus becomes a model dependent
statement. Using uncorrelated, independent equation of state estimates, we
propose a model independent figure-of-merit. Our approach involves the inverse
of the sum of inverse variances of uncorrelated $w(z_i)$ bins as an way to
capture all of the available information related to dark energy in a given data
set or experiment.

The paper is organized as follows: In the next Section we review techniques for
reconstructing the EOS, we following the methods of Ref.~\cite{HutCoo} and
Ref.~\cite{Rie06}. In Section~3 we present our results, addressing whether dark
energy is a cosmological constant or has dynamical behavior that leads to an
evolution in the EOS. We show that existing data rule out extreme forms of
dynamical behavior at the 68\% confidence level, including models that either
start with $w(z) > -0.4$ at $z>1.0$ or models that asymptote to values of $w(z)
> -0.85$ at $z< 0.2$. In Section~3.1 we outline a new figure of merit to assess
the dark energy information content of future experiments, based on uncorrelated
binned estimates of the EOS. We conclude with a summary of results in Section~4.

\begin{figure}[t]
\centerline{\psfig{file=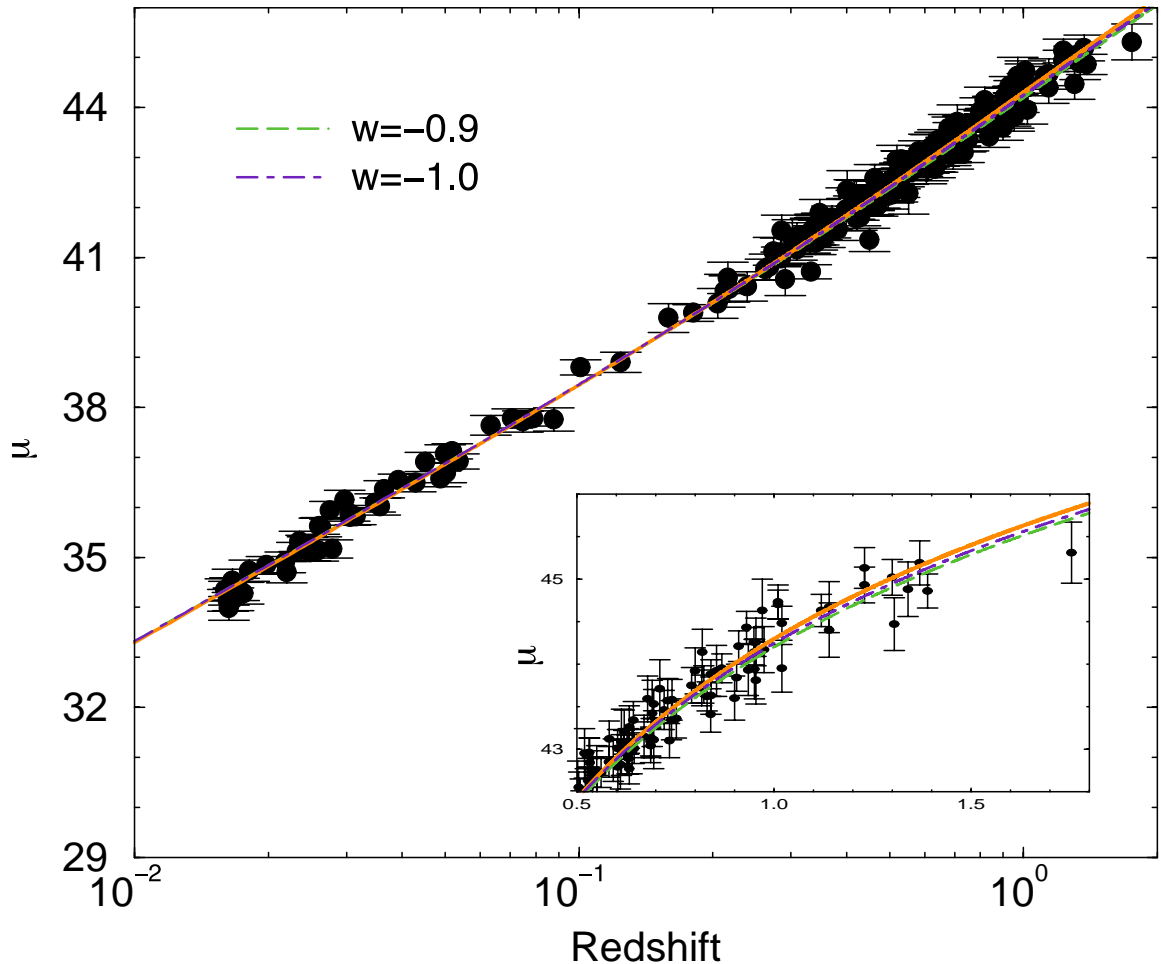,width=6.0in,angle=0}}
\caption{Hubble diagram for type Ia supernova data used for the present
analysis. The dataset includes a total of 192 SNe from the recent 
analysis of light curves in Ref.~\cite{Davis07}. For comparison, we also
separately analyze a subset of 104 SNe within this sample that were also used in
Riess et al.~\cite{Rie06}. The distance moduli used here for this subset,
however, are different from this original study due to the reanalysis of light
curves in Davis et al. \cite{Davis07}. The orange line is for our best fit.}
\label{fig:data}
\end{figure}

\section{Methodology}
\label{sec:methods}
A simple way to model the dark component of the universe credited for the
accelerating expansion is through a modification of the standard cosmological
model. We utilize the Friedmann equations, and specify the dark energy density
and its equation-of-state (EOS). We assume a piecewise constant EOS, with value
$w_i$ in each $i$\textsuperscript{th} redshift bin (defined by an upper boundary
at $z_i$).  We fit the observational data to the luminosity distance as a
function of redshift.  The expression for luminosity distance, $d_L(z)$, depends
on whether the universe is flat, positively, or negatively curved (i.e., the
sign of $\Omega_k$), and is given by
\begin{eqnarray}
\label{eq:DL}
d_L(z) = (1+z) \frac{c}{H_0} 
 &\times& \left\{ \begin{array}{l} \frac{1}{\sqrt{|\Omega_k|}}   {\rm
sinh}{\left(\sqrt{|\Omega_k|}\int_0^z \frac{dz'}{E(z')}\right)} \; \quad \quad
{\rm if} \; \Omega_k > 0, \\ 
\int_0^z \frac{dz'}{E(z')}  \; \quad \quad \quad \quad \quad \quad \quad \quad
\quad \quad {\rm if} \; \Omega_k = 0,                  \\
\frac{1}{\sqrt{|\Omega_k|}} \sin{\left(\sqrt{|\Omega_k|}\int_0^z
\frac{dz'}{E(z')}\right)} \; \; \quad \quad {\rm if} \;  \Omega_k < 0,
\end{array}\right.
\end{eqnarray}
where
\begin{equation}
E(z)= \Big[ \Omega_M (1+z)^3 + \Omega_k (1+z)^2+ (1-\Omega_k-\Omega_M) F(z) \Big]^{\frac{1}{2}} \, ,
\end{equation}
and $F(z)$ depends on the binning of $w(z)$. For the $n$\textsuperscript{th}
redshift bin $F(z)$ has the form
\begin{equation}
F(z_n>z>z_{n-1})= (1+z)^{3(1+w_n)}\prod_{i=0}^{n-1}(1+z_i)^{3(w_i-w_{i+1})}.
\end{equation}
We define the zeroth bin as $z_0 = 0$, so the product is unity for redshift $z$
in the first bin.  For our primary analysis we set $z_1 = 0.2$, $z_2 = 0.5$,
$z_3 = 1.8$, and $z_4$ extends beyond the surface of last scattering at $z_{CMB}
= 1089$.  We assume $w(z>1.8) = -1$ and allow variation within the remaining
three redshift bins.  Selecting the cutoff point for $z_3$ is fairly arbitrary;
we found that pushing it back as far as $z_3 = 2.5$ does not substantially
alter the outcome of our analysis.

In addition to SNe, we also make use of four primary constraints from the
literature following the analysis of Riess et al.\cite{Rie06}, modifying these
to account for variations in curvature since the analysis in Ref.~\cite{Rie06}
assumed an {\em a priori} flat cosmological model. These constraints are:
\begin{itemize}
\item The product of the Hubble parameter $ h
\equiv {H_0}/{100}$ and the present local mass density $\Omega_m$ from SDSS
large scale structure measurements \cite{TegSDSS}, given by $ \Omega_m h = 0.213 \pm
0.023$. In cases where we allow curvature to vary, we either take a flat, broad
prior in curvature or, to highlight the result under the assumption of a measured
value for the curvature, we take $\Omega_k = -0.014 \pm 0.017$ as derived by
{\it WMAP} analysis by combining {\it WMAP} and the Hubble constant \cite{Spergel06}.

\item The SDSS luminous red galaxy baryon acoustic oscillation (BAO) distance
estimate to redshift $z_{\rm BAO}=0.35$.  Here the constraint is on the overall
parameter $A \equiv \frac{\sqrt{\Omega_M H_0^2}}{c z_{\rm BAO}} \left[r^2(z_{\rm
BAO}) \frac{cz_{\rm BAO}}{H_0 E(z_{\rm BAO})}\right]^{1/3}$, where
$r(z)=d_L(z)/(1+z)$ is the angular diameter distance.  The angular correlation
function of red galaxies in the SDSS spectroscopy survey leads to $A=
0.469(\frac{n}{0.98})^{-0.35} \pm 0.017$ \cite{Eis05}.  Following Riess et
al. \cite{Rie06}, we use the WMAP estimate for the scalar tilt with $n=0.95$~\cite{Spergel06}.

\item The distance to last scattering, at $z_{\rm CMB}=1089$, written in the
dimensionless form $R_{CMB} \equiv \frac{\sqrt{\Omega_M H_0^2}}{c} r(z_{\rm
CMB})$ where $r(z_{\rm CMB})$ is the angular diameter distance to the CMB last
scattering surface.  We use the dark energy and curvature independent estimate
with $R=1.70 \pm 0.03$ \cite{Wang07}.

\item The distance ratio between $z=0.35$ and last scattering $z=1089$ as measured by the SDSS BAO analysis \cite{Eis05}:
\begin{equation}
R_{0.35} = \frac{\left[r^2(z_{\rm BAO}) \frac{cz_{\rm BAO}}{H_0 E(z_{\rm BAO})}\right]^{1/3}}{r(z_{\rm CMB})},
\end{equation}
with the value of $R_{0.35}=0.0979 \pm 0.0036$.
\end{itemize}

When estimating parameters we make use of the $\chi^2$ statistic for a
particular model with parameter set $\theta$ ($w_i$, $H_0$, $\Omega_m$, and in some
cases $\Omega_k$):
\begin{equation}
\chi^2(\theta) = \sum_{n=1}^N \left(\frac{\mu^{{\rm theory}}_i - \mu^{{\rm data}}_i}{\sigma^2_i + \sigma^2_{{\rm int}}}\right),
\end{equation}
where $N$ is the total number of supernovae in the sample. While our total
sample includes 192 supernovae, we also extract a subset of 104 supernovae that
was analyzed previously \cite{Rie06}. While the subsample is for comparison with
previous results, our distance estimates differ from the original analysis due
to a reanalysis of light curves by Davis et al. \cite{Davis07} using a common
light curve fitting method. When estimating $\chi^2$ we set an intrinsic
dispersion of order $\sigma_{{\rm int}}
\sim 0.1$, such that the minimum $\chi^2$ value for the best fit model comes to
be about one. Using $\chi^2$ we calculate the probability $P(\theta)
\propto \exp(-\frac{\chi^2(\theta)}{2})$ and derive constraints by generating
Markov chains through a Monte-Carlo algorithm.  The algorithm generates a set of
models whose members appear in the set (or chain) a number of times proportional
to their likelihood of being a good fit to the observed data, after
marginalizing over other priors.  The likelihood probability functions for
each independent parameter are generated by simply taking a histogram over the
chain. We marginalize over $H_0$ assuming a broad uniform prior over the range
$[30,85]$ km s$^{-1}$ Mpc$^{-1}$. We also marginalize over $\Omega_m$ assuming
the quoted prior above for $\Omega_mh$ using SDSS large scale structure
measurements.

In the case of dark energy parameters, the redshift binned EOS estimates are
correlated such that their errors depend upon each other.  These correlations in
the redshift binned EOS estimates, $w_i$, are captured by the covariance matrix,
and this matrix can be generated by taking the average over the chain such that:
\begin{equation}
C = \left\langle {\bf w} {\bf w}^T \right\rangle - \left\langle {\bf w}\right\rangle \left\langle {\bf w}^T \right\rangle.
\end{equation}
This covariance matrix is not diagonal as the values found for the various EOS
estimates are correlated.  This is expected, as the integration over low
redshift bins in Eq.~(\ref{eq:DL}) obviously affects the model fit in middle and
higher redshift bins.  The behavior is also such that the best constraints are
found for the lowest redshift bin, with higher bins having progressively weaker
constraints. With the addition of the distance scale to the last scattering
surface from CMB data, the constraint in the higher redshift bins are improved,
as seen in prior analyses \cite{Rie06}.

Instead of discussing $w(z)$ in correlated bins, we follow Huterer \& Cooray
\cite{HutCoo} and transform the covariance matrix to decorrelate the EOS estimates.
This is achieved by changing the basis through an orthogonal matrix rotation that diagonalizes the covariance matrix.
We start by the definition of the Fisher matrix,
\begin{equation}
{\bf F} \equiv C^{-1} = {\bf O^T} \Lambda {\bf O}
\end{equation}
where the matrix $\Lambda$ is the diagonalized covariance for transformed bins.
The uncorrelated parameters are then defined by the rotation performed by the orthogonal rotation matrix ${\bf q} = {\bf O} {\bf w}$
 and has the covariance matrix of $\Lambda^{-1}$.
There is freedom of choice in what orthogonal matrix is used to perform this transformation, and we use the
particular choice that was advocated in Ref.~\cite{HutCoo} and write the weight transformation matrix:
\begin{equation}
\tilde{W} = {\bf O}^T \Lambda^{\frac{1}{2}} {\bf O}
\label{eq:transmatrix}
\end{equation}
where the rows are then summed such that the weights from each band add up to
unity.  This choice ensures we have mostly positive contributions across all
bands, an intuitively pleasing result, so for example we can interpret the
weighting matrix as an indication of how much a measurement in the third bin is
influenced by SNe in the first and second bins.  We apply the transformation
$\tilde{W}$ to each link in the Markov chain to generate a set of independent,
uncorrelated measures of the EOS and its probability distributions as determined
by the observables.

In addition to probability distribution functions for each of the uncorrelated
EOS estimates, to study the redshift evolution of $w(z)$, we also study the
differences between these uncorrelated estimates. The probability distribution
functions of the difference between estimate $w_i$ and $w_j$ is generated
through $P(w_{\rm diff}) \propto \int P_{w_i}(w)P_{w_j}(w+w_{\rm diff}) dw$.
While in the uncorrelated case an estimate $w_i$ is not necessarily associated
with a single redshift bin, due to support from adjacent redshift bins due to
the transformation, any significant difference between $w_i$ and $w_j$ can be
considered to be evidence for an evolution in the dark energy EOS. Moreover, if
the dark energy is associated with a cosmological constant, then these
difference estimates should be precisely zero.

\begin{figure}[t]
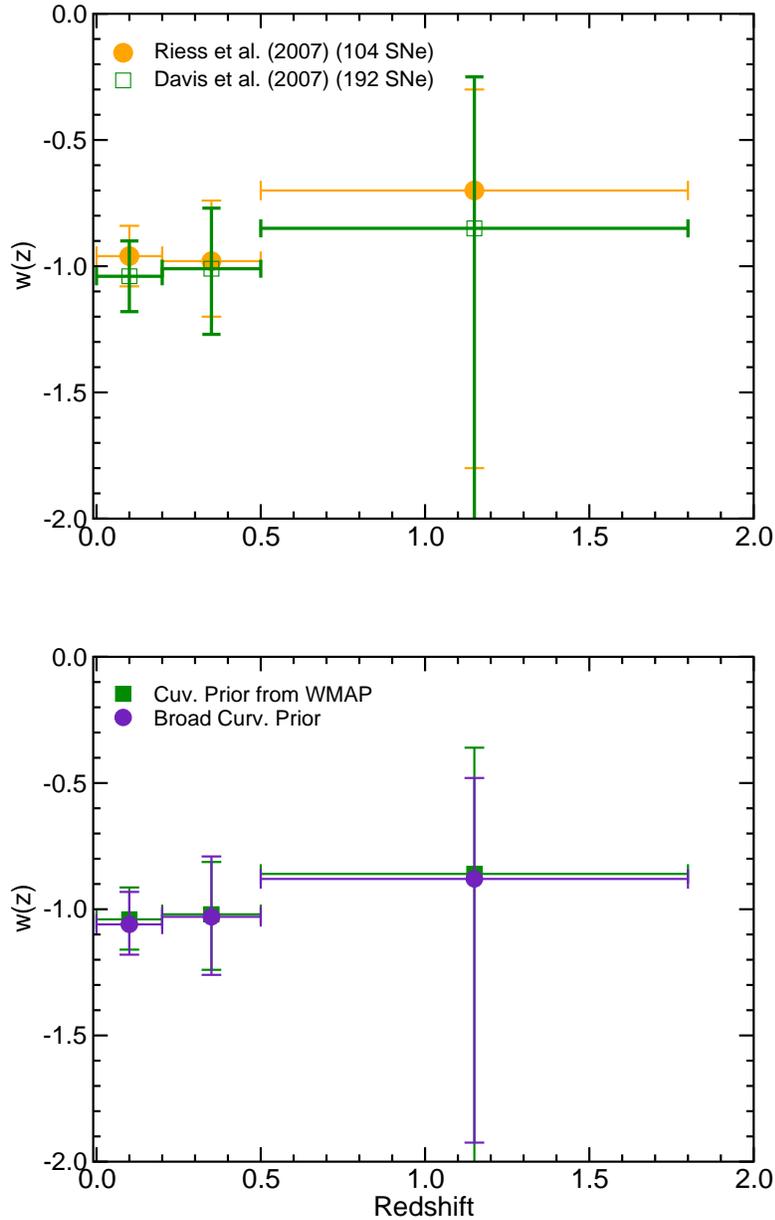

\centerline{\psfig{file=wzflat.eps,width=4.0in,angle=0}}
\vspace{0.9cm}
\centerline{\psfig{file=wzcurv.eps,width=4.0in,angle=0}}
\caption{Uncorrelated estimates of the dark energy equation of state using a
combined sample of supernova data and constraints from {\it WMAP} and BAO
measurements. In the top panel a flat universe prior is assumed
($\Omega_k=0$), and the filled and open symbols show the constraint with the
total sample of 192 SNe from Ref.~\cite{Davis07} and a subset of 104 SNe
corresponding to a previous analysis \cite{Rie06}, respectively.  In the bottom
panel we show $w(z)$ estimates without the flat universe prior; open
and filled symbols showing constraints with a broad prior for $\Omega_k$ and
$\Omega_k=-0.014 \pm 0.017$, respectively.}
\label{fig:wz}
\end{figure}

\begin{figure}[t]
\centerline{\psfig{file=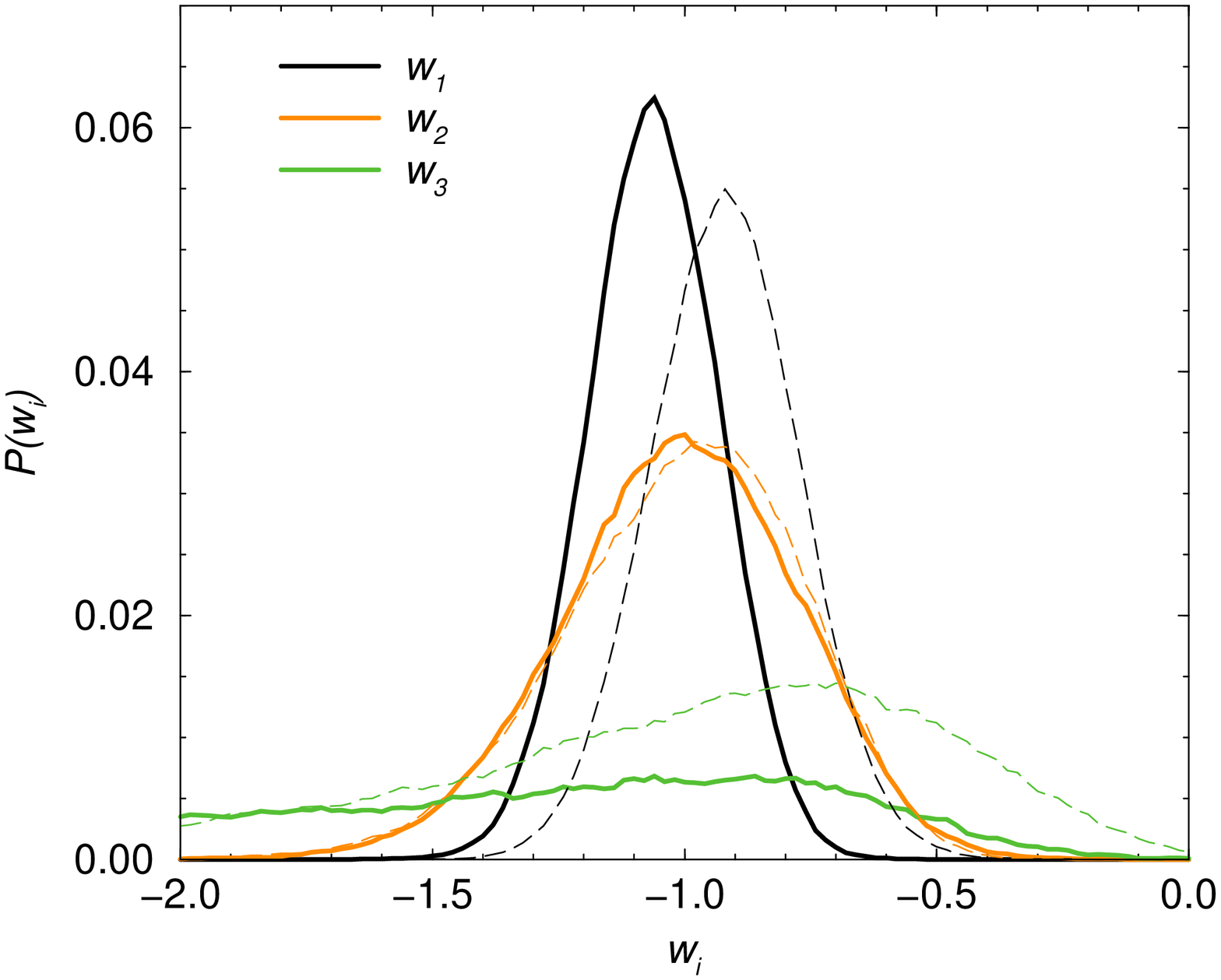,width=4.0in,angle=-90}}
\centerline{\psfig{file=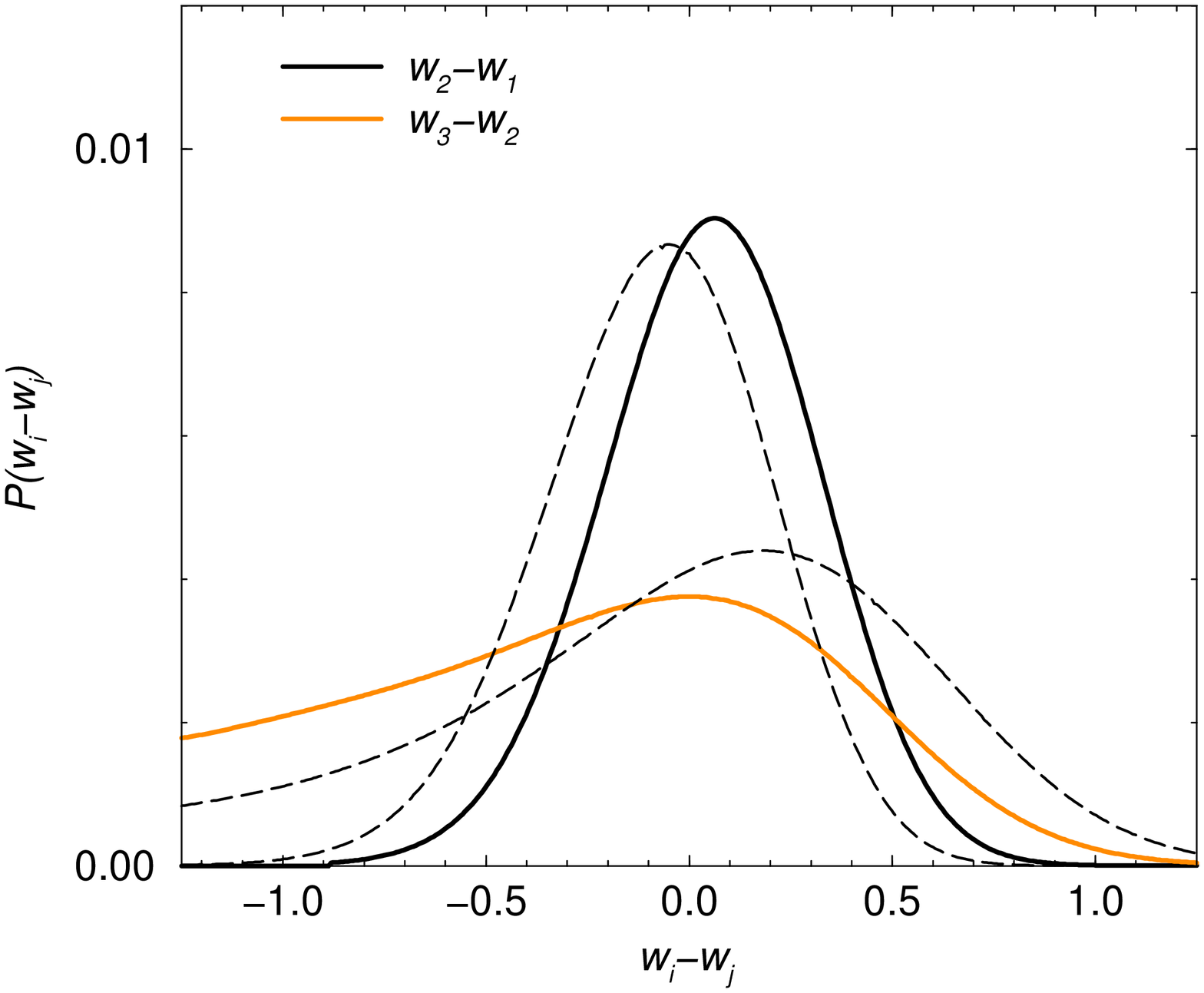,width=4.0in,angle=-90}}
\caption{Probability distribution functions of
uncorrelated estimates of the dark energy equation of state (top panel) and the
difference between estimates of the equation of state (bottom panel). Solid and
dashed lines are for the case with 192 and 104 SNe (see caption of Figure~1),
respectively.}
\label{fig:prob}
\end{figure}

\begin{figure}[t]
\centerline{\psfig{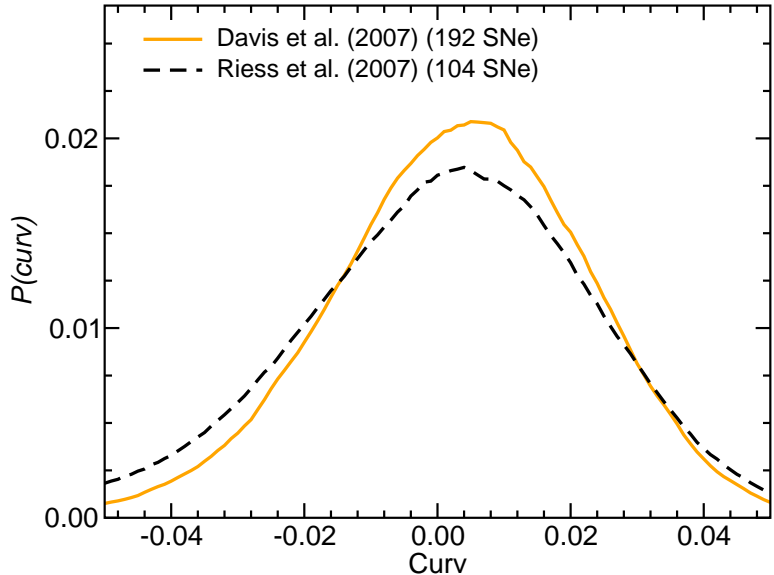}}
\caption{
Probability distribution function of cosmic curvature by combining supernova
data with additional constraints outlined in the paper. We assume a broad, flat
prior for $\Omega_k$, but the combination of SN data and CMB and BAO distance
scales results in a tight constraint on curvature. We find $\Omega_k=0.004 \pm
0.019$ (192 SNe) and $\Omega_k=0.002 \pm 0.022$ (104 SNe), which is fully
consistent with estimates made by the {\it WMAP} analysis by combining {\it
WMAP} data with supernova data from the SNLS survey.}
\label{fig:pcurv}
\end{figure}

\begin{figure}[t]
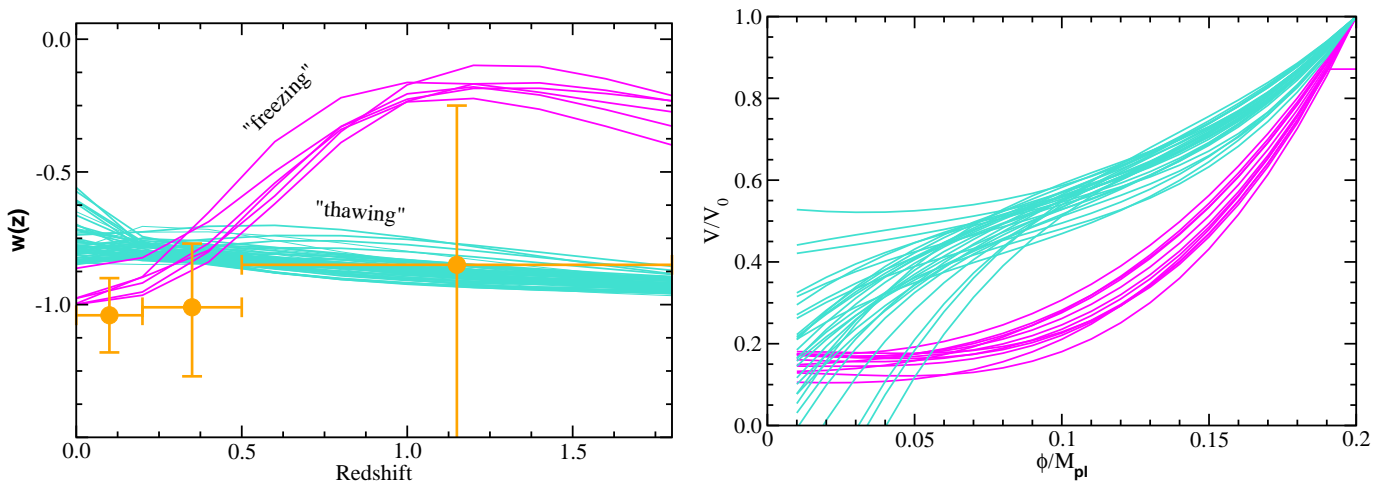

\centerline{\psfig{file=Models.eps,width=3.5in,angle=0}
\hspace{0.1cm}
\psfig{file=PotentialsEssence2.eps,width=3.5in,angle=0}}
\caption{{\it Left}: A comparison of $w(z)$ estimates and dynamical dark energy
models, based on the Monte-Carlo modeling approach of Ref.~\cite{HutPei},
that are inconsistent with current estimates. We show both cases of ``thawing'' and ``freezing'' models inconsistent with current
data (see text for details). {\it Right:} The shapes of potentials generally corresponding to $w(z)$ models shown in
the left panel which are inconsistent with our estimates of the binned EOS from a combined sample of supernovae and cosmological
distance scale measurements. We show $V(\phi)$ as a function of the scalar field $\phi$ when
potentials are normalized to the value at $z=3$ ($V_0 \equiv V(z=3)$), which we take to be $\phi=0.2$.}
\label{fig:wzmodels}
\end{figure}

\begin{figure}[t]
\centerline{\psfig{file=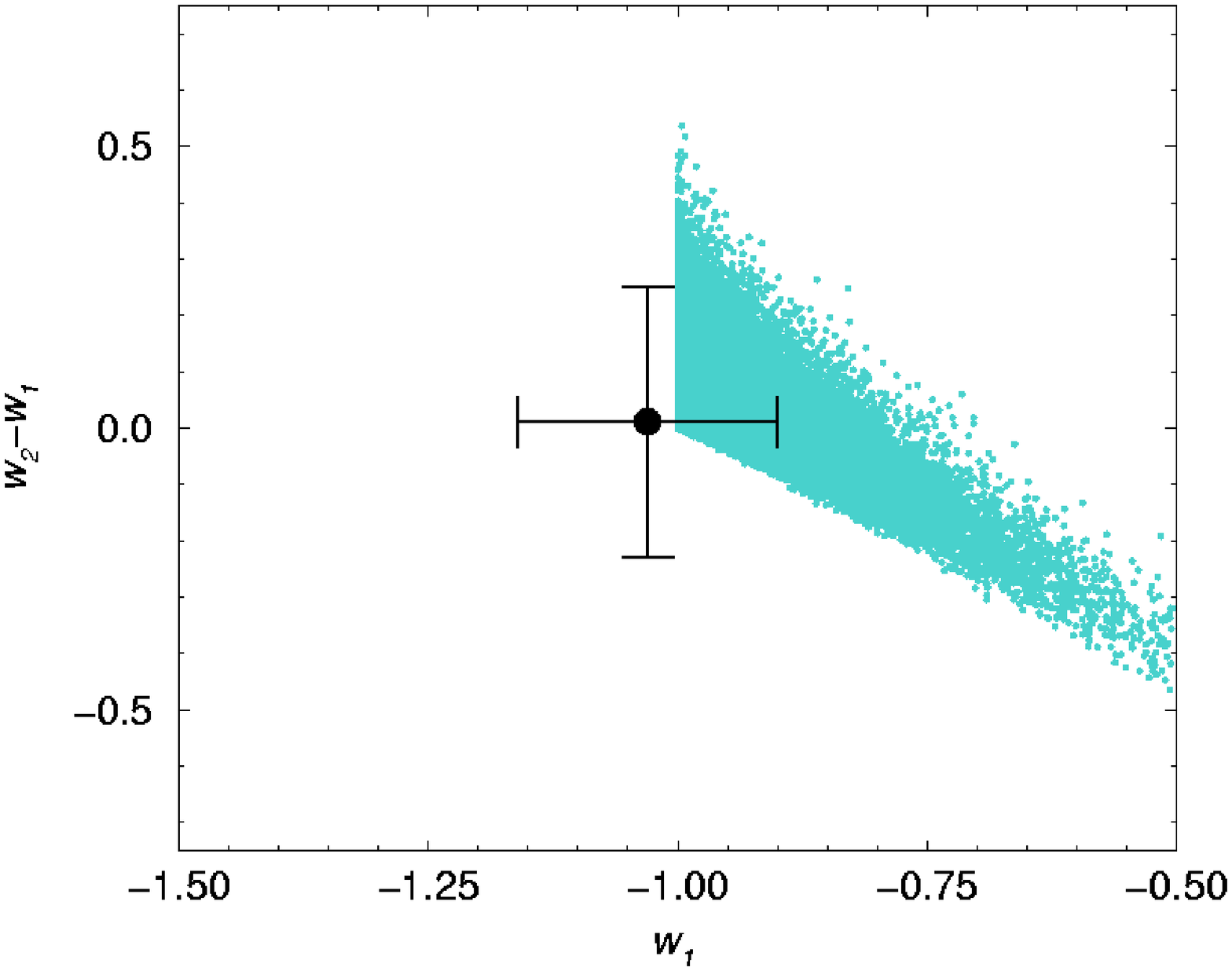,width=4.0in,angle=-90}}
\centerline{\psfig{file=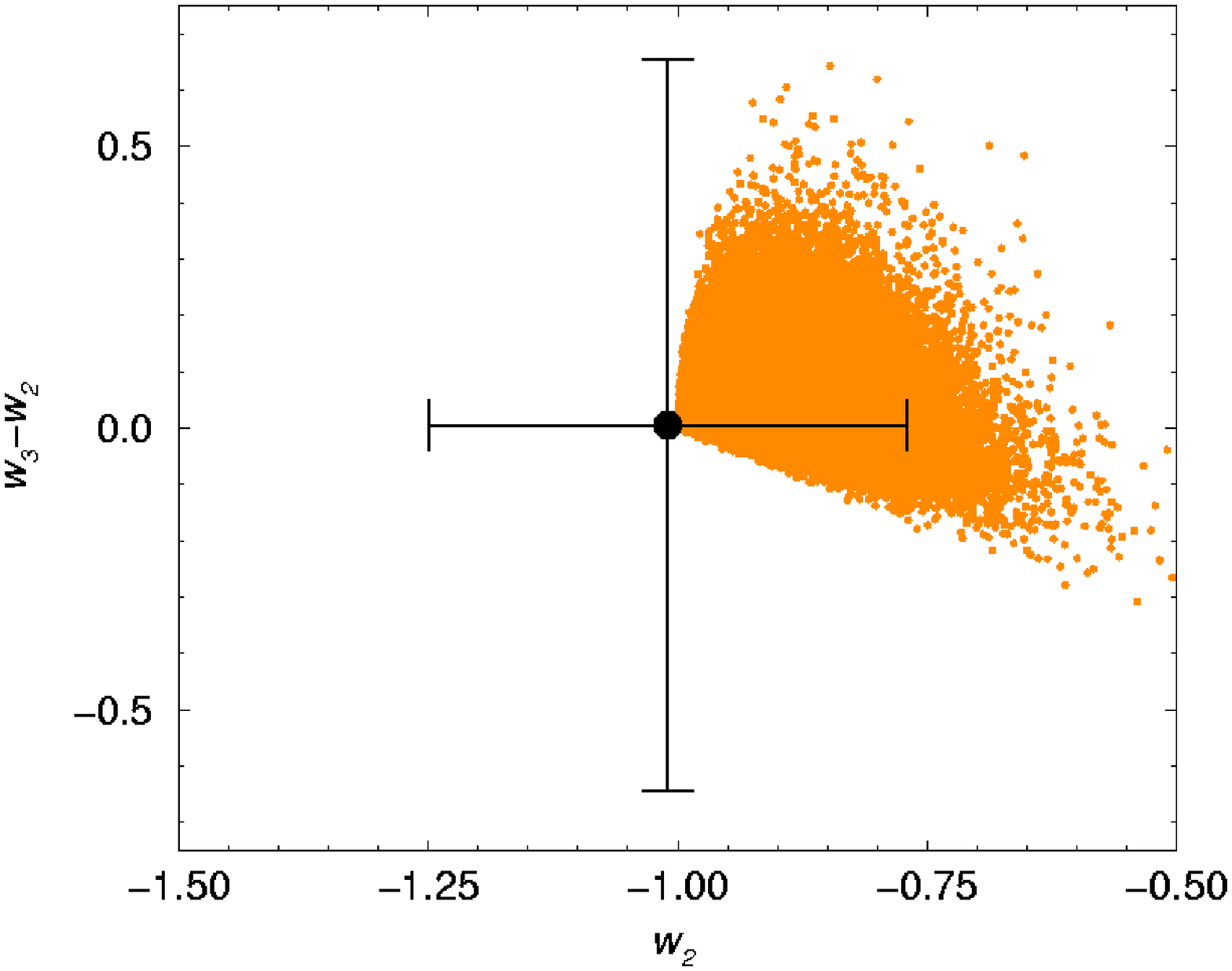,width=4.0in,angle=-90}}
\caption{A comparison of $w(z)$ values from dynamical dark
energy models, based on the Monte-Carlo modeling approach of Ref.~\cite{HutPei}. Each dot
represents a potential dynamical dark energy model. The data points are the
allowed range from our analysis.}
\label{fig:wzdiff}
\end{figure}

\section{Results}

The results presented in this paper are derived from the statistical analysis of
a combination of recent supernova surveys, including the Supernova Legacy Survey
(SNLS) \cite{Astier}, the ESSENCE survey \cite{Wood}, and high-$z$ supernovae
discovered by the Hubble Space Telescope (HST) \cite{Rie06}.  In particular, we
use a total of 192 SNe Ia measurements taken from a combination of supernovae
analyzed in Ref.~\cite{Davis07} using a common light curve fitting method
(Figure~1).  Also, for comparison, results are presented for
the 104 SNe Ia that overlap with the ``Gold'' data set presented in Riess et
al. \cite{Rie06}, although the distance moduli values we use here for the same
subsample are slightly different from the values published in the original
analysis due to variations in the light curve fitting. This analysis
includes the four external constraints outlined in the previous
section. In addition to the standard flat cosmological model generally assumed
when making fits to dark energy EOS, we also allow for variations in the
curvature, both with a prior on $\Omega_k$ based on {\it WMAP} and Hubble
constant measurements, and with no prior.

In Figure~2 we highlight our results for $w(z)$, in the redshift bins $z<0.2$,
$0.2 < z <0.5$ and $1.8>z>0.5$, with $w(z>1.8) = -1$.  These binned estimates
are uncorrelated following the technique of Huterer \& Cooray
\cite{HutCoo} and as outlined in Section~II.  The procedure to decorrelate
binned estimates that are predefined over a certain range usually results in
adding contributions from nearby bins, but these are generally smaller than the
main contribution from the bin in which the estimate was first defined.  On
average, the first uncorrelated measure is $73\%$ determined by its own bin,
with a minimal contribution from the third bin.  The second measure, on average,
was $50\%$ determined by the second bin, with a substantial contribution from
the first bin.  The third measure was typically $54\%$ determined by the third
bin, with a $6\%$ contribution from the first bin.

These results do not presume a particular evolutionary history, as opposed to
model fitting to a specific form, e.g. $w(z)=w_0+w_1 z$ \cite{CooHut99} or
$w(z)=w_0+w_a(1-a)$. Our procedure, which fits binned values of $w(z_i)$ and
then decorrelates them, has the advantage that one can extract redshift
evolution independent of a model. This is particularly effective if the model to
be assumed turns out not to be an accurate representation of the true underlying
EOS.

As shown in Figure~2, the dark energy EOS as a function of redshift is fully
consistent with $w(z)=-1$ at the 1$\sigma$ confidence level, for both the full
sample of 192 SNe, and the subset of 104 SNe corresponding to the earlier
analysis~\cite{Rie06}. As shown in the lower panel of Figure~2, this conclusion
is unchanged when we drop the assumption related to a flat cosmological model,
regardless of the assumed prior on $\Omega_k$. Our external constraints related
to distance to the last scattering surface from CMB, and BAO distance scale to
$z=0.35$, provide a strong constraint on $\Omega_k$. We explicitly include an
additional flatness constraint to allow comparison with earlier work
\cite{Davis07,Rie06}.  To highlight the extent to which EOS estimates are
consistent with $w(z)=-1$, in Figure~3 we plot the probability distribution
functions $P(w_i)$ both for flat cosmologies and for the case with curvature
allowed to vary.  Except for the third bin, which still remains mostly undefined
with a very broad probability function, the first two bins are peaked and allow
constraints at a high confidence level over $-2 < w < 0$. These show a clear
consistency with $w=-1$, and hence are completely consistent with a cosmological
constant. While the allowed range for $w(z_3)$ is broad, at the 68\% confidence
level we find that $w(z_3) < -0.2$, suggesting that we can rule out a large EOS
even at $z>0.5$.

The uncorrelated binned estimates of $w(z)$ derived in Huterer \& Cooray
\cite{HutCoo} using an earlier ``Gold'' sample from Riess et al. \cite{Rie04}
showed an equation of state that varied significantly between the
lowest-redshift bin and the second bin. This difference decreased in the most
recent ``Gold'' sample as analyzed by Riess et al. \cite{Rie06}. In the current
work, utilizing an extended sample of supernovae, we no longer find evidence for
a variation in the dark energy EOS between the first and the second bin. To show
this explicitly, we also plot the probability distribution function of the
difference between binned estimates of $w_i(z)$ in Figure~3.  Between the first
and the second bin we find the difference to be $w_2-w_1=0.06 \pm 0.26$ at the
68\% confidence level for the 192 SNe sample with a flat model.  Current data is
thus completely consistent with a cosmological constant. Previous estimates of a
large and a statistically significant value for $w_2-w_1$, with $w_1 < -1$ and
$w_2 > -0.8$, led to suggestions in the literature for a physical mechanism
called dark energy ``metamorphosis''
\cite{Alam}. While earlier conclusions were limited to a small set of supernovae
data, with the larger sample it is now clear that there is little evidence for a
sudden transition in the EOS around $z \sim 0.2$. Future data could tighten
these constraints, either further narrowing down to a cosmological constant, or
providing evidence for small variations in the EOS with redshift.

We show that our conclusions are generally unchanged by assumptions related to
the curvature. This is because we constrain the EOS using a combination of
supernova data and existing measurements of the cosmic distance scale out to
$z=1089$ and $z=0.35$ with CMB and BAO, respectively. The
combination of supernova data and these measurements, combined with our prior
on the Hubble constant, leads to a strong independent constraint on the
curvature parameter $\Omega_k$. We show the probability distribution
$P(\Omega_k)$ in Figure~4 for both the full sample and the subset of 104
supernovae. In both cases $\Omega_k$ is consistent with zero; with the full
supernovae sample, we find $\Omega_k=0.004 \pm 0.019$. This is
about 1$\sigma$ away from the combined {\it WMAP} and Hubble constant estimate of
$\Omega_k= -0.014 \pm 0.017$ or the combined {\it WMAP}+SNLS estimate of
$\Omega_k=-0.011 \pm 0.012$ \cite{Spergel06}. On the other hand, the combined
{\it WMAP}+SDSS estimate from the same analysis is
$\Omega_K=-0.0053^{+0.0068}_{-0.0060}$, which is a shift in the direction of the
$\Omega_k$ value we find when the combined SNe dataset is analyzed with the {\it WMAP}
and BAO priors.

To demonstrate how our estimates of $w(z_i)$ can be used to understand the redshift
evolution of the dark energy component, in Figure~5 we show a sample of
predictions related to dynamical dark energy models that are ruled out with
present estimates of $w_i(z)$ at the 68\% confidence level. These
cases generally involve a dark energy EOS that starts as $w > -0.2$ at high
redshifts, or an EOS that stars with a value around -1 at high redshift but evolve
to a value greater than -0.85 when $z<0.2$. The former models belong to the
general category of ``freezing'' models described in Caldwell \& Linder
\cite{Caldwell}, while the latter models are categorized as ``thawing''
models. We generate these models following the numerical technique 
of Huterer \& Peiris \cite{HutPei}, writing the Klein-Gordon equation for the
evolution related to a scalar field as $\ddot{\phi} + 3H\dot{\phi} + dV/d\phi=0$,
and then numerically generating a large number of models following the
Monte-Carlo flow approach as used for numerical models of inflation
\cite{Kinney,Liddle,PeiEas,Efs,FriCoo}. We do not reproduce details as the process is similar to
the modeling of Ref.~\cite{HutPei,Chon}.

As a further application of our estimates of the EOS, and the difference between
two binned estimates of EOS, in Figure~6 we compare our measured value with
values expected for a large number of models. Again, we rule out certain extreme
models where $w(z_i)$ varies rapidly between adjacent bins.  In dynamical dark
energy models, models that generally lead to a large variation in $w$ between
two adjacent bins also have a value significantly different from -1 in one of
the bins. Thus most models are currently ruled out from the value of $w$ in a
single bin, rather than through the difference of $w_2-w_1$ or $w_3-w_2$, since
the the latter are still largely uncertain.  While we can use the numerical
modeling technique of Huterer \& Peiris~\cite{HutPei} to make qualitative
statements about the EOS, and to rule out extreme possibilities for its
dynamical evolution, given the stochastic nature of model generation we cannot
use this sort of method to make detailed statements about, for example, the
scalar field (quintessence) potential responsible for dark energy. Instead, it
is necessary to directly reconstruct the scalar field potential from supernovae
distance data.  While there are attempts to recover the potential by directly
model fitting various parametric forms of the potential as a function of the
scalar field, such as power-law or polynomial functions of the scalar field
$\phi$, model independent binned estimates of the potential are
preferable~\cite{Li}.

\subsection{A New Figure of Merit}

As discussed in the previous sections, our binned estimates allow us to study
the redshift evolution of the dark energy EOS without the need to assume an
underlying model. This is to be contrasted with the usual approach, in which a
parameterized form for $w(z)$ is required to fit the data.  With an increasing
supernova sample size, and improvement in other cosmological observations, we
may be able to recover three or more binned values at the 10\% level or better.
In the context of planning dark energy experiments, and assessing the
constraining power of future data, it may be advantageous to consider binned
estimates of the dark energy EOS, rather than the error associated with a
specific and arbitrary two parameter model of the equation of state.  The latter
is the approach adopted to quote the ``figure of merit'' (FOM) of an experiment,
based on the inverse of the area of the ellipse of the two parameters describing
the EOS with redshift (introduced in~\cite{DETF}). In the case of binned
estimates, once uncorrelated, we can quote an alternative FOM as $\left[\sum_i
1/\sigma^2(w_{z_i})\right]^{1/2}$, which takes into account the combined inverse
variance of all independent estimates of the EOS.  In an upcoming paper we will
quantify the exact number of $w(z)$ estimates that can be determined with future
experiments involving supernovae and large-scale structure (weak lensing, baryon
acoustic oscillations over a wide range in redshift), and we will compare this
alternative FOM to the method of Ref.~\cite{DETF}.

\section{Summary}

We use a sample of 192 SNe Ia (and a subset for comparison) to constrain the
dark energy equation-of-state parameter and its variation as a function of
redshift.  We use a model independent approach, providing uncorrelated
measurements across three redshift bins below $z=1.8$, and find that $w(z)$ is
consistent with a cosmological constant ($w(z)=-1$) at the $68\%$ confidence level. At
the same confidence level we find that the EOS is greater than $-0.2$ over the
redshift range of $0<z<1.8$.  Overall, there is no strong evidence
against the assumption of a flat $\Omega_k = 0$ universe, especially when recent
supernova data are combined with cosmological distance scale measurements from
{\it WMAP} and BAO experiments.  We argue against previous claims in the literature of
evolving dark energy, such as dark energy ``metamorphosis'', where the EOS
changes significantly from $w > -1$ at $z > 0.2$ to $w<-1$ at $z<0.2$.  Instead,
we find consistency with a cosmological constant, encapsulated in the 68\% level
constraint: $w_2-w_1=0.06 \pm 0.26$, where $w_1$ is the value of the dark energy
EOS in the $z<0.2$ bin, and $w_2$ is the value in the bin $0.2<z<0.5$.
A transition in the EOS can also be ruled out between our second and third
binned estimates of EOS, although we still find large uncertainties in our
determination of the EOS at $z>0.5$, and we are insensitive to rapid variations at
$z>1$.  We compare our EOS estimates to Monte Carlo generated dynamical dark
energy models associated with a single scalar field potential. Our EOS estimates
generally allow us to rule out extreme ``thawing'' and ``freezing'' models,
though a large number of potential shapes remain in agreement with current data.

We also suggest an alternative, parameter independent figure-of-merit, with
which to evaluate the potential of future missions to constrain properties of
the dark energy.

\section{Acknowledgments}
AC and DEH are partially supported by the DOE at LANL
through IGPP grant Astro-1603-07. AC acknowledges funding for undergraduate research,
at UC Irvine as part of NSF Career AST-0645427, which was used to support SS,
and thanks UCI Undergraduate Research Opportunities
Program (UROP) for additional support in the form of a UCI Chancellor's Award for Fostering of
Distinguished  Undergraduate Research. DEH acknowledges a
Richard P. Feynman Fellowship from LANL.


\section*{References}

\begin{thebibliography}{99}
\frenchspacing

\bibitem{Rieetal04}
A.~G.~Riess {\it et al.},
B.~J.~Barris {\it et al.},
arXiv:astro-ph/0310843; R.~A.~Knop {\it et al.},
arXiv:astro-ph/0309368;
J.~L.~Tonry {\it et al.},
Astrophys.\ J.\  {\bf 594}, 1 (2003).

\bibitem{perlmutter-1999}
        S.\ Perlmutter {\it et al.},  Astrophys. J. {\bf 517}, 565 (1999);
        A.\ Riess {\it et al.}, Astron. J. {\bf 116}, 1009 (1998).

\bibitem{Rie06}
  A.~G.~Riess {\it et al.},
  arXiv:astro-ph/0611572.

\bibitem{Wood}
 W.~M.~Wood-Vasey {\it et al.},
  arXiv:astro-ph/0701041.


\bibitem{Davis07}
  T.~M.~Davis {\it et al.},
  arXiv:astro-ph/0701510.


\bibitem{Spergel}
  D.~N.~Spergel {\it et al.}  [WMAP Collaboration],
  Astrophys.\ J.\ Suppl.\  {\bf 148}, 175 (2003);

\bibitem{Spergel06}
  D.~N.~Spergel {\it et al.}  [WMAP Collaboration],
  arXiv:astro-ph/0603449.

\bibitem{Padmanabhan}
  T.~Padmanabhan,
  Phys.\ Rept.\  {\bf 380}, 235 (2003)
  [arXiv:hep-th/0212290].

\bibitem{Astier} P. Astier et al., Astron. Astrophys. {\bf 447} 31 (2006), astro-ph/0510447.

\bibitem{Caldwell98}
  R.~R.~Caldwell, R.~Dave and P.~J.~Steinhardt,
  Phys.\ Rev.\ Lett.\  {\bf 80}, 1582 (1998)
  [arXiv:astro-ph/9708069].

\bibitem{Garnavich}
  P.~M.~Garnavich {\it et al.}  [Supernova Search Team Collaboration],
  Astrophys.\ J.\  {\bf 509}, 74 (1998)
  [arXiv:astro-ph/9806396].

\bibitem{Visser}
  M.~Visser,
  Class.\ Quant.\ Grav.\  {\bf 21}, 2603 (2004)
  [arXiv:gr-qc/0309109].

\bibitem{HT99} D. Huterer and M. S. Turner, \prd {\bf 60} 081301
(1999).

\bibitem{Sahlen}
  M.~Sahlen, A.~R.~Liddle and D.~Parkinson,
  Phys.\ Rev.\ D {\bf 72}, 083511 (2005)
  [arXiv:astro-ph/0506696].



\bibitem{Li07}
  C.~Li, D.~E.~Holz and A.~Cooray,
  Phys.\ Rev.\  D {\bf 75}, 103503 (2007)
  [arXiv:astro-ph/0611093].


\bibitem{CooHut99}
  A.~R.~Cooray and D.~Huterer,
Astrophys.\ J.\  {\bf 513}, L95 (1999) [arXiv:astro-ph/9901097].

\bibitem{Lin03}
  E.V. Linder, \PRL\ {\bf 90} 091301 (2003).

\bibitem{Gerke}
  B.~F.~Gerke and G.~Efstathiou,
   ``Probing quintessence: Reconstruction and parameter estimation from
  Mon.\ Not.\ Roy.\ Astron.\ Soc.\  {\bf 335}, 33 (2002)
  [arXiv:astro-ph/0201336].


\bibitem{HutSta03}
D.~Huterer and G.~Starkman,
Phys.\ Rev.\ Lett.\  {\bf 90}, 031301 (2003).

\bibitem{HutCoo}
  D.~Huterer and A.~Cooray,
  Phys.\ Rev.\ D {\bf 71}, 023506 (2005)
  [arXiv:astro-ph/0404062].

\bibitem{Caldwell}
  R.~R.~Caldwell and E.~V.~Linder,
  Phys.\ Rev.\ Lett.\  {\bf 95}, 141301 (2005)
  [arXiv:astro-ph/0505494].

\bibitem{Wang}
  Y.~Wang and M.~Tegmark,
  Phys.\ Rev.\ D {\bf 71}, 103513 (2005)
  [arXiv:astro-ph/0501351].

\bibitem{DETF}
  A.~Albrecht {\it et al.},
  arXiv:astro-ph/0609591.

\bibitem{Teg04}
  M.~Tegmark {\it et al.}  [SDSS Collaboration],
  Phys.\ Rev.\  D {\bf 69}, 103501 (2004)
  [arXiv:astro-ph/0310723].

\bibitem{Rie05}
  A.~G.~Riess {\it et al.},
  Astrophys.\ J.\  {\bf 627}, 579 (2005)
  [arXiv:astro-ph/0503159].


\bibitem{Eis05}
  D.~J.~Eisenstein {\it et al.}  [SDSS Collaboration],
  Astrophys.\ J.\  {\bf 633}, 560 (2005)
  [arXiv:astro-ph/0501171].

\bibitem{Wang07}
  Y.~Wang and P.~Mukherjee,
  arXiv:astro-ph/0703780.


\bibitem{HutPei}
  D.~Huterer and H.~V.~Peiris,
  arXiv:astro-ph/0610427.


\bibitem{Rie04}
  A.~G.~Riess {\it et al.}  [Supernova Search Team Collaboration],
  Astrophys.\ J.\  {\bf 607}, 665 (2004)
  [arXiv:astro-ph/0402512].

\bibitem{Alam}
  U.~Alam, V.~Sahni, T.~D.~Saini and A.~A.~Starobinsky,
  Mon.\ Not.\ Roy.\ Astron.\ Soc.\  {\bf 354}, 275 (2004)
  [arXiv:astro-ph/0311364].

\bibitem{Kinney}
  W.~H.~Kinney,
  Phys.\ Rev.\  D {\bf 66}, 083508 (2002)
  [arXiv:astro-ph/0206032].

\bibitem{Liddle}
  A.~R.~Liddle,
  Phys.\ Rev.\  D {\bf 68}, 103504 (2003)
  [arXiv:astro-ph/0307286].

\bibitem{PeiEas}
  H.~Peiris and R.~Easther,
  JCAP {\bf 0607}, 002 (2006)
  [arXiv:astro-ph/0603587].

\bibitem{Efs}
  S.~Chongchitnan and G.~Efstathiou,
  Phys.\ Rev.\  D {\bf 72}, 083520 (2005)
  [arXiv:astro-ph/0508355].

\bibitem{FriCoo}
  B.~C.~Friedman, A.~Cooray and A.~Melchiorri,
  Phys.\ Rev.\  D {\bf 74}, 123509 (2006)
  [arXiv:astro-ph/0610220].

\bibitem{Chon}
  S.~Chongchitnan and G.~Efstathiou,
  arXiv:0705.1955 [astro-ph].

\bibitem{Li}
  C.~Li, D.~E.~Holz and A.~Cooray,
  Phys.\ Rev.\  D {\bf 75}, 103503 (2007)
  [arXiv:astro-ph/0611093].
  
\bibitem{TegSDSS}
  M.~Tegmark {\it et al.}  [SDSS Collaboration],
  Astrophys.\ J.\  {\bf 606}, 702 (2004)
  [arXiv:astro-ph/0310725].


\end{thebibliography}
\end{document}